\documentclass[fleqn]{cas-sc}

\usepackage{amsmath}
\usepackage{amssymb}
\usepackage{bm}
\usepackage{graphicx}
\usepackage{xcolor}
\usepackage[authoryear,longnamesfirst]{natbib}

\definecolor{bleudefrance}{rgb}{0.19, 0.55, 0.91}

\def\refeq#1{(\ref{#1})}
\def\reffig#1{Fig.~\ref{#1}}
\def\reftab#1{Tab.~\ref{#1}}

\def\t{\bm{t}^\alpha}
\def\tt{\hat{\bm{t}}^\alpha}

\def\n{\bm{n}^\alpha}
\def\nn{\hat{\bm{n}}^\alpha}
\def\nnn{\hat{\bm{n}}'^\alpha}

\def\m{\bm{m}^\alpha}
\def\mm{\hat{\bm{m}}^\alpha}
\def\NN{\bm{N}^\alpha}
\def\MM{\bm{M}^\alpha}
\def\NNN{\bm{N}'^\alpha}

\def\gplane#1{\{#1\}}
\def\gdir#1{\langle#1\rangle}
\def\new#1{#1}

\newcommand\systab[4]{#1 & (#2)[#3] & [#3] & [#2] & [#4]}

\setcitestyle{round}

\begin{document}
\let\WriteBookmarks\relax
\def\floatpagepagefraction{1}
\def\textpagefraction{.001}

\title[mode = title]{Symmetry-adapted single crystal yield criterion for non-Schmid materials}
\shorttitle{Symmetry-adapted single crystal yield criterion for non-Schmid materials}
\shortauthors{R. Gr\"oger}

\author[1]{Roman Gr\"oger}[orcid=0000-0002-7116-2774]
\ead{groger@ipm.cz}
\cortext[cor1]{Corresponding author}
\address[1]{Institute of Physics of Materials and CEITEC IPM, Czech Academy of Sciences, \v{Z}i\v{z}kova 22, Brno 61600, Czech Republic}
\cormark[1]

\begin{abstract}[S U M M A R Y]
All yield criteria that determine the onset of plastic deformation in crystalline materials must be invariant under the inversion symmetry associated with a simultaneous change of sign of the slip direction and the slip plane normal. We demonstrate the consequences of this symmetry on the functional form of the effective stress, where only the lowest order terms that obey this symmetry are retained. A particular form of \new{yield criterion} is obtained for materials that do not obey the Schmid law\new{, hereafter called non-Schmid materials.} \new{Application of this model} to body-centered cubic and hexagonal close-packed metals shows under which conditions the non-Schmid stress terms become significant in predicting the onset of yielding. In the special case, where the contributions of all non-Schmid stresses vanish, this model reduces to the maximum shear stress theory of Tresca.
\end{abstract}


\maketitle


\section{Introduction}

Yield criteria play indispensable roles in identifying the stress states that cause the initiation of elastic-plastic deformation \citep{hill:98}. \new{At the single-crystal level, the onset of plastic deformation is associated with the stress state that activates dislocation glide on the most highly stressed slip system compatible with the underlying crystallographic space group. \citet{schmid:50} postulated that slip on a particular crystallographic system commences when the shear stress parallel to the slip direction acting in the slip plane reaches its critical value. This argument constitutes a central point in the maximum shear stress theory developed already by \citet{tresca:1864}.} \new{Almost a century ago, Taylor and Elam \citep{taylor:26, taylor:28} pointed out that single crystals of $\alpha$-Fe and $\beta$-CuZn do not behave according to the rules established by the Schmid law. Since then, the pool of materials for which the Schmid law is not capable of correctly predicting the initiation of plastic flow grew steadily to include body-centered cubic (bcc) metals \citep{christian:83, duesbery:84, ito:01}, B2 and L1$_2$ intermetallics \citep{paidar:84}, \mbox{CuZnAl} \citep{alkan:18}, single crystal Ni-base superalloys \citep{tinga:10}, NiTi shape memory alloys \citep{alkan:17} \new{and Inconel 718 steel \citep{ghorbanpour:17}}. It is now widely accepted that the main distinguishing factor between the yielding of close-packed and non-close-packed metals are non-planar cores of screw dislocations in the latter structures \citep{ito:01, bassani:01, vitek:08}. Moreover, the widespread observation of the breakdown of the Schmid law implies that Tresca's model of yielding is a special case of a more general yield criterion.}

\begin{figure}
  \centering
  \includegraphics[scale=0.8]{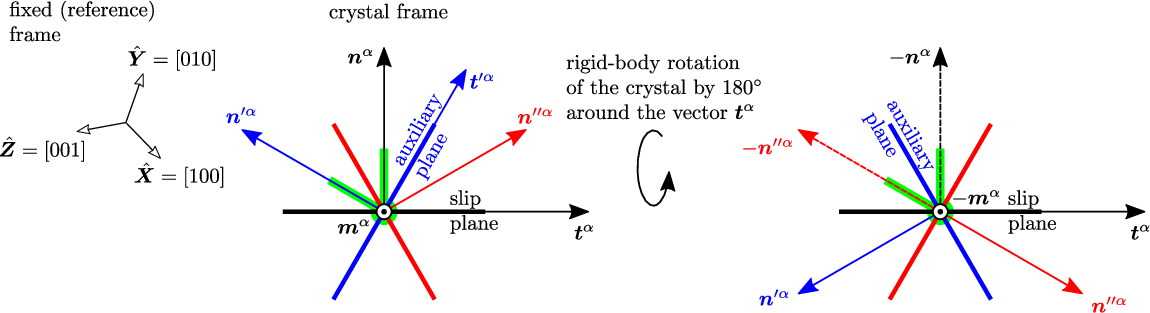}
  \caption{Schematic illustration of the symmetry relations \refeq{eq:symm} and \refeq{eq:symm2} by rigid-body rotations about the axis $\t$. The auxiliary system as defined in the text is drawn in blue. The effective yield criterion $\tau^*(\m,\n,\bm{n}'^\alpha|\bm{\sigma})$ is initially described in terms of the lattice vectors $\m$, $\n$ and $\bm{n}'^\alpha$ which are highlighted in green. After rotation of the crystal by 180$^\circ$ about the axis $\t$, the same green motif refers to the lattice vectors $-\m$, $-\n$, $-\bm{n}''^\alpha$ and the corresponding effective stress is $\tau^{*\alpha}(-\m,-\n,-\bm{n}''^\alpha|\bm{\sigma})$. Both values of $\tau^{*\alpha}$ must be the same, which proves the validity of the condition \refeq{eq:symm2}.}
  \label{fig:rotsym}
\end{figure}

\new{The importance of non-Schmid stresses was first recognized by \citet{asaro:77} on the basis of their \new{finite deformation studies of strain localization} in ductile crystals. They observed the formation of shear bands that resulted from stress bifurcations, which were attributed to micromechanical processes such as cross-slip of screw dislocations. If the crystal deforms by multiple slip, lattice rotation plays an important role in the localization process \citep{dao:93}. In order to quantify the effect of non-Schmid stresses on the initiation of elastic-plastic deformation, \citet{qin:92} proposed a simple yield criterion}
\begin{equation}
  \tau^{*\alpha}(\m,\n|\bm{\sigma}) \leq \tau^*_{cr} \ ,
  \label{eq:yieldcrit}
\end{equation}
where the effective stress $\tau^{*\alpha}$, corresponding to the slip system $\alpha$, was written as a linear combination of several components of the stress tensor $\bm{\sigma}$ with coefficients representing the relative importance of each stress. The work of \citet{duesbery:84}, \citet{ito:01}, as well as our more recent contribution \citep{groger:14a, groger:19a}, have demonstrated that the relevant stress components are primarily shear stresses acting both parallel and perpendicular to the slip direction. \new{To incorporate these effects, the effective stress assigned to the slip system $\alpha$ was expressed using stress components resolved in two orthogonal systems mutually rotated by $60^\circ$ in the zone of the slip direction $\m$ (see \reffig{fig:rotsym}a). The slip plane in the first (reference) system has normal $\n$, whereas the so-called non-glide plane in the second (auxiliary) system has normal $\bm{n}'^\alpha$.} In terms of the stresses resolved in these systems, the effective stress was defined as
\begin{equation}
  \tau^{*\alpha} = \sigma_{nm}^\alpha + a_1\sigma_{n'm}^\alpha + a_2\sigma_{tn}^\alpha + a_3\sigma_{t'n'}^\alpha \ ,
  \label{eq:taueff_current}
\end{equation}
where $a_1$, $a_2$, $a_3$ are non-Schmid coefficients that can be determined from atomistic simulations or from experiments. \new{Here, $\sigma_{nm}=\bm{\sigma}:(\nn\otimes\mm)$ and $\sigma_{n'm}=\bm{\sigma}:(\nnn\otimes\mm)$ are shear stresses parallel to the slip direction, and $\sigma_{tn}=\bm{\sigma}:[(\nn\times\mm)\otimes\nn]$, $\sigma_{t'n'}=\bm{\sigma}:[(\nnn\times\mm)\otimes\nnn]$ shear stresses perpendicular to the slip direction, all resolved in the black and blue coordinate frames in \reffig{fig:rotsym}(a). Hereafter, overhat represents a unit vector, colon the double-dot product (which yields a scalar), and the symbol $\otimes$ implies the tensor/element-wise product.}

So far, these yield criteria have been developed for the bcc metals Mo \citep{groger:08a, lim:13, daphalapurkar:18}, W \citep{groger:08a, lim:13, cereceda:16, kraych:19}, Ta \citep{lim:13, alleman:14, cho:18}, ferromagnetic $\alpha$-Fe \citep{chen:13, lim:15}, non-magnetic phase of Cr \citep{groger:20}, hexagonal close-packed (hcp) Mg \citep{ostapovets:18, qiu:21}, as well as for the shape memory alloys NiTi \citep{alkan:17}, CuZnAl \citep{alkan:18}\new{, and Inconel 718 steel \citep{ghorbanpour:17}}. Similar formulations of the yield criteria that go beyond the Schmid law were proposed independently by other authors \citep{yalcinkaya:08, koester:12}. The orientational dependence of the critical resolved shear stress \new{(CRSS)} for all bcc metals was characterized also in DFT simulations by \citet{dezerald:15a}. These atomic-level details associated with the glide of isolated $1/2\gdir{111}$ screw dislocations in bcc metals were introduced into mathematical models \citep{caillard:03, dorn:64, groger:21} that provide the activation enthalpy to transform the dislocation into its critical state under the applied stress \citep{groger:08c, proville:13, dezerald:15, pi:17}. Although non-Schmid effects in bcc metals originate from non-planar cores of $1/2\gdir{111}$ screw dislocations, their effect was also observed in interacting dislocation networks studied using discrete dislocation dynamics models \citep{srivastava:13, weygand:15}. The yield criteria involving non-Schmid stress terms form the basis of constitutive equations used in a number of crystal plasticity models \citep{knezevic:14, patra:14, cereceda:16, keshavarz:16, savage:17, mapar:17, cho:18, savage:18, zecevic:18, daphalapurkar:18}, and in kinetic Monte Carlo calculations \citep{stukowski:15}. Computational aspects of this new theoretical framework were studied for both single crystals \citep{steinmann:98, bassani:11} and for random polycrystals \citep{racherla:07, groger:08b, bassani:11}. The mathematical basis of these models in the framework of finite deformations was investigated by \citet{cleja-tigoiu:13}, and \citet{soare:14}. Despite the other developments in the field, \citet{le:13} have argued that a properly regularized Schmid law can accommodate the observed variations of the critical resolved shear stress. However, this suggestion is at odds with a more recent phonon stability analysis of \citet{salahshoor:18}, which demonstrates that the onset of plastic deformation is dominated by non-Schmid effects arising mostly from short wavelength instabilities. Most recently, \citet{pal:21} observed non-Schmid behavior in molecular dynamics simulations of cyclotrimethylene trinitramine ($\beta$-HMX) subjected to hydrostatic pressures up to 27 GPa.

The effective yield criterion should not only capture the results of atomistic simulations and experiments, but the symmetry of the effective stress must be compatible with crystallography. To be more specific, consider an arbitrary slip system $\alpha$ in bcc crystals that is characterized by two vectors: $\n$ representing the normal of the slip plane, and $\m$ that is parallel to the slip direction. \new{The two possible choices for the orientation of the normal of the slip plane imply that the Burgers circuit can be made around the \emph{same} dislocation in two \emph{opposite} directions.} For example, the $(\bar{1}01)[111]$ system can be described by two pairs of vectors $(\m,\n)$: (i) $\m=[111]$ and $\n=[\bar{1}01]$, or (ii) $\m=[\bar{1}\bar{1}\bar{1}]$ and $\n=[10\bar{1}]$. \new{These two choices lead to  opposite slip directions $\pm\m$ and thus to opposite directions of the Burgers vector of the dislocation, $\pm\bm{b}=\pm\mm b$, where $b$ is the magnitude of the Burgers vector, and $\mm$ is a unit vector parallel to the slip direction. However, \emph{the Burgers vector is a topological quantity and does not depend on the choice of the coordinate frame}. In the spirit of Neumann's principle \citep{neumann:1885}, we thus demand that any physical quantity involving the slip system $\alpha$ must be invariant under a \emph{simultaneous} change of sign of the vectors $\m$ and $\n$.} Therefore, the effective stresses involved in all single crystal yield criteria must obey the inversion symmetry
\begin{equation}
  \tau^{*\alpha}(\m,\n|\bm{\sigma}) = \tau^{*\alpha}(-\m,-\n|\bm{\sigma}) \ .
  \label{eq:symm}
\end{equation}


\new{It can be easily shown that Eq.~\refeq{eq:symm} is a statement of invariance of $\tau^{*\alpha}$ under rigid-body rotation. In \reffig{fig:rotsym}, we define the orientation of an arbitrary slip system $\alpha$ by three lattice vectors $\t$, $\n$, $\m$, all written relative to the fixed frame. Assume that the effective stress for this system is $\tau^{*\alpha}(\m,\n|\bm{\sigma})$. Now, carry out a rigid-body rotation of the crystal by 180$^\circ$ around the axis $\t$, which brings $\m\rightarrow-\m$ and $\n\rightarrow-\n$.  For this new orientation, the effective stress is $\tau^{*\alpha}(-\m,-\n|\bm{\sigma})$. Obviously, in non-polar crystal structures, rigid-body rotations by $180^\circ$ are symmetry operations and all physical quantities must be invariant under this transformation, which proves the validity of Eq.~\refeq{eq:symm}. It is evident from \refeq{eq:taueff_current} that only the Schmid stress $\sigma_{nm}$ in the first term obeys the symmetry \refeq{eq:symm}. However, this symmetry is broken by the shear stress perpendicular to the slip direction, $\sigma_{tn}$, which changes sign under the transformation $(\m,\n) \rightarrow (-\m,-\n)$. The symmetry \refeq{eq:symm} is broken also for the remaining two stress terms, but this matter requires a deeper analysis that will be made in the next Section.}

Despite the fact that the yield criterion \refeq{eq:yieldcrit} with the effective stress defined by \refeq{eq:taueff_current} incorporates correctly the effect of non-Schmid stresses, it does not obey the fundamental inversion symmetry \refeq{eq:symm} that is dictated by crystallography. \new{When using these yield criteria, only certain combinations of ($\m$,$\n$) are allowed for which the coefficients of the yield criterion were determined from atomistic simulations or from experiments. However, the equivalent pairs $(-\m,-\n)$ must be avoided, because they would result in different $\tau^{*\alpha}$ for crystallographically identical systems.} To remove this drawback \new{and to avoid ambiguity when predicting the onset of slip activity}, we first derive a general form of the effective stress as a power series expansion in terms of all components of the stress tensor and investigate which stress terms are allowed by the symmetry \refeq{eq:symm}. The obtained yield criterion is then applied to materials of cubic and hexagonal symmetry to predict the primary and secondary slip systems for a few combinations of their parameters.


\section{Symmetry-adapted effective stress}
\label{sec:theory}

\new{The stress state at a point of an elastic body is characterized by six components of a symmetric stress tensor. An infinite number of possibilities exist to combine these stress components to arrive at a scalar quantity (hereafter called as an effective stress), which can be used in the yield criteria to determine whether the material is locally in the state of purely elastic or elastic-plastic deformation. The simplest representation of this effective stress, due to Tresca, is based on the validity of the Schmid law. In the following, our objective
is to generalize this formulation to all crystalline materials in which the onset of plastic deformation depends also on other (non-Schmid) stresses.}

\new{\subsection{Nonlinear effective stress}}

\new{
We begin by writing the effective stress ($\tau^*$) as a weighted linear combination of six stress functions $f_{ij}(\sigma_{ij})$,
\begin{equation}
  \tau^* = \sum_{i=1}^3 \sum_{j=i}^3 c_{ij} f_{ij}(\sigma_{ij}) \ .
  \label{eq:taueff0}
\end{equation}
Each function $f_{ij}(\sigma_{ij})$ will be defined by its Maclaurin series as $f_{ij}(\sigma_{ij}) = \sum_{n=0}^\infty (1/n!)\sigma_{ij}^nf_{ij}^{(n)}(0)$, where the superscript $(n)$ represents the $n$-th derivative with respect to $\sigma_{ij}$. To simplify the notation, we introduce a new set of coefficients $c_{ijn} = c_{ij}(1/n!)f_{ij}^{(n)}(0)$. Incorporating the last two expressions into \refeq{eq:taueff0} then yields an equivalent representation of the effective stress,
\begin{equation}
  \tau^* = \sum_{i=1}^3\sum_{j=i}^3 (c_{ij0} + c_{ij1}\sigma_{ij} + c_{ij2}\sigma_{ij}^2 + \dots)\ ,
  \label{eq:taueff}
\end{equation}
where $ij=\{tt,tn,tm,nn,nm,mm\}$ refer to the three axes in \reffig{fig:rotsym}. We first require that $\tau^* = 0$ for zero applied stress, i.e. when all $\sigma_{ij}=0$. This can be satisfied easily by setting all $c_{ij0}$ to zero.}

\new{In order to write \refeq{eq:taueff} for a particular slip system $\alpha$, we first resolve all stress components $\sigma_{ij}^\alpha$ using the unit vectors of the slip plane normal $\nn$ and the slip direction $\mm$ corresponding to the system $\alpha$. The vector orthogonal to both is then $\tt = \nn \times \mm$. Only the components of $\sigma_{ij}^\alpha$ that obey the inversion symmetry defined by Eq.~\refeq{eq:symm} will be retained. All three normal stresses, i.e. $\sigma_{tt}^\alpha = \bm{\sigma} : (\tt \otimes \tt)$, $\sigma_{nn}^\alpha = \bm{\sigma} : (\nn \otimes \nn)$, and $\sigma_{mm}^\alpha = \bm{\sigma} : (\mm \otimes \mm)$, and also the shear stress $\sigma_{nm}^\alpha = \bm{\sigma} : (\nn \otimes \mm)$ are clearly invariant under this symmetry. To obtain the simplest possible form of the effective stress for the system $\alpha$, we retain in \refeq{eq:taueff} only the lowest-order (i.e., linear) terms containing these stresses. The remaining two shear stresses perpendicular to the slip direction are defined as $\sigma_{tm}^\alpha = \bm{\sigma} : (\tt \otimes \mm)$ and $\sigma_{tn}^\alpha = \bm{\sigma} : (\tt \otimes \nn)$. Both these stresses violate the aforementioned inversion symmetry in that they change sign as $(\mm,\nn) \rightarrow (-\mm,-\nn)$. Therefore, in order to comply with \refeq{eq:symm}, only even powers of these stresses are allowed in the yield criterion with the lowest order being quadratic. Owing to these arguments, it follows from Eq.~\refeq{eq:taueff} that the effective stress exerted on any slip system $\alpha$ that satisfies the condition \refeq{eq:symm} has the following form:}
\begin{equation} 
  \tau^{*\alpha} = c_{tt1}\sigma_{tt}^\alpha + c_{tn2}(\sigma_{tn}^\alpha)^2 + c_{tm2}(\sigma_{tm}^\alpha)^2 +
  c_{nn1}\sigma_{nn}^\alpha + c_{nm1}\sigma_{nm}^\alpha +
  c_{mm1}\sigma_{mm}^\alpha \ .
  \label{eq:taueff2}
\end{equation}
If all non-Schmid contributions to $\tau^{*\alpha}$ vanish (i.e., when $c_{tt1}=c_{tn2}=c_{tm2}=c_{nn1}=c_{mm1}=0$), $\tau^{*\alpha}$ must equal to the Schmid stress $\sigma_{nm}^\alpha$. \new{One possibility\footnote{Another possibility is to divide both sides of \refeq{eq:taueff2} by $c_{nm1}$. This would renormalize all remaining coefficients in the right-hand side of \refeq{eq:taueff2} as well as the effective yield stress $\tau^*_{cr}$. \new{However, the form of the effective stress would remain the same.}} to satisfy this condition is to take $c_{nm1}=1$.} In the other extreme, i.e. when all non-Schmid stresses play role in the initiation of plastic deformation, it is customary to express $\tau^{*\alpha}$ in terms of the hydrostatic stress $\sigma_h^\alpha$. The presence of $\sigma_h^\alpha$ and $\sigma_{mm}^\alpha$ causes that the two remaining normal stresses ($\sigma_{tt}^\alpha$ and $\sigma_{nn}^\alpha$) are not independent. Starting with \refeq{eq:taueff2}, it is possible to rewrite the first, the fourth, and the sixth term using $\sigma_{nn}^\alpha-\sigma_{tt}^\alpha$, $\sigma_h^\alpha$ and $\sigma_{mm}^\alpha$. This results in an equivalent but physically more transparent representation of the effective stress
\begin{equation} 
  \tau^{*\alpha} =  \sigma_{nm}^\alpha + c_{tm2} (\sigma_{tm}^\alpha)^2 + a_2 (\sigma_{nn}^\alpha-\sigma_{tt}^\alpha) + c_{tn2} (\sigma_{tn}^\alpha)^2 +
  a_4 \sigma_{mm}^\alpha + a_5\sigma_h^\alpha \ ,
  \label{eq:taueff2_general}
\end{equation}
where we have introduced new coefficients to replace the linear combinations of $c_{ijn}$. The first term in \refeq{eq:taueff2_general} is the resolved Schmid stress for the system $\alpha$. The second term is the shear stress parallel to the slip direction acting in the plane with the normal $\t$ (this is perpendicular to the slip direction and lies in the slip plane). The third term is a combination of two normal stresses, which generates a \emph{shear} stress parallel to the slip direction in another coordinate system, where the axes $\t$ and $\n$ are rotated by $45^\circ$ in the zone of $\m$. The fourth term is directly the shear stress perpendicular to the slip direction. The fifth and the sixth terms incorporate the effects of the normal stress parallel to the dislocation line and of the hydrostatic stress, respectively. A notable feature of \refeq{eq:taueff2_general} is that all stress components entering $\tau^{*\alpha}$ are resolved in a \emph{single} coordinate system spanned by the vectors $\t$, $\n$ and $\m$.

\new{\subsection{Linear effective stress using an auxiliary system}}

Although the form of the effective stress \refeq{eq:taueff2_general} obeys the inversion symmetry \refeq{eq:symm}, the presence of quadratic terms may lead to a number of complications in homogenizations and developments of plastic flow rules for random and textured polycrystals \citep{dao:93, bassani:94}. In the following, we demonstrate that a form equivalent to \refeq{eq:taueff2_general} can be obtained if the power series in \refeq{eq:taueff} is terminated \emph{before} the quadratic terms $(\sigma_{tm}^\alpha)^2$ and $(\sigma_{tn}^\alpha)^2$. Since the linear terms of these stresses are absent by symmetry, the effective stress reduces to a simple form
\begin{equation} 
  \tau^{*\alpha} \approx \sigma_{nm}^\alpha + a_2 (\sigma_{nn}^\alpha-\sigma_{tt}^\alpha) + a_4 \sigma_{mm}^\alpha + a_5\sigma_h^\alpha \ .
  \label{eq:taueff3_simple}
\end{equation}
All stresses in this expression are resolved in a single orientation defined by the vectors $\t$, $\n$ and $\m$. Therefore, this form cannot capture the variation of the CRSS with the orientation of the maximum resolved shear stress plane (MRSSP) and on the orientation of the shear stress perpendicular to the slip direction, as explained in the Introduction. The simplest way to re-introduce these effects is to write $\tau^{*\alpha}$ as a linear superposition of the stresses $\sigma_{nm}^\alpha$ and $\sigma_{nn}^\alpha-\sigma_{tt}^\alpha$ resolved in \emph{two different} orientations in the zone of the slip direction, which is the same step as taken previously by \citet{qin:92} to define their effective stress \refeq{eq:taueff_current}. This superposition extends \refeq{eq:taueff3_simple} to
\begin{equation} 
  \tau^{*\alpha} =  \sigma_{nm}^\alpha + a_1\sigma_{n'm}^\alpha + a_2 (\sigma_{nn}^\alpha-\sigma_{tt}^\alpha) + a_3 (\sigma_{n'n'}^\alpha-\sigma_{t't'}^\alpha) + a_4 \sigma_{mm}^\alpha + a_5\sigma_h^\alpha \ ,
  \label{eq:taueff3}
\end{equation}
where we have added the terms with coefficients $a_1$ and $a_3$, in which $\bm{n}'^\alpha$ is the normal of an auxiliary plane in the zone of $\m$. The explicit forms of the individual stress components in \refeq{eq:taueff3} are given in \reftab{tab:sigcomp} (column ``bcc crystals''). The presence of $\bm{n}'^\alpha$ and $\bm{t}'^\alpha$ requires generalization of the symmetry condition \refeq{eq:symm}. We can see from \reffig{fig:rotsym} that a rigid-body rotation of the crystal by $180^\circ$ about the axis $\t$ results in the transformations $\m \rightarrow -\m$, $\n \rightarrow -\n$ and $\bm{n}'^\alpha \rightarrow -\bm{n}''^\alpha$. The last transformation involves normals of two different planes, which is a natural consequence of introducing the auxiliary plane. In order for the effective stress \refeq{eq:taueff3} to be invariant under this transformation, we thus require that it  satisfies the symmetry condition
\begin{equation}
  \tau^{*\alpha}(\m,\n,\bm{n}'^\alpha|\bm{\sigma}) = \tau^{*\alpha}(-\m,-\n,-\bm{n}''^\alpha|\bm{\sigma}) \ .
  \label{eq:symm2}
\end{equation}
Similarly as Eq.~\refeq{eq:symm} defines the symmetry of the quadratic form of $\tau^{*\alpha}$ in \refeq{eq:taueff2_general}, Eq.~\refeq{eq:symm2} defines the symmetry of the physically more appealing linear form of $\tau^{*\alpha}$ in \refeq{eq:taueff3}. It is demonstrated in \reffig{fig:rotsym} that both these relations are statements of physical invariance under rigid-body rotations about the axis $\t$.

\new{In the previously used effective stress \refeq{eq:taueff_current}, the effects of shear stresses perpendicular to the slip direction are incorporated by directly including the \emph{shear stresses}  $\sigma_{tn}^\alpha$ and $\sigma_{t'n'}^\alpha$. We have shown that these stress terms violate the symmetry represented by \refeq{eq:symm} and \refeq{eq:symm2}. On the contrary, in the effective stress \refeq{eq:taueff3}, the effects of these shear stresses are incorporated via \emph{differences of normal stresses} $\sigma_{nn}-\sigma_{tt}$ and $\sigma_{n'n'}-\sigma_{t't'}$. These stresses do not violate the aforementioned symmetries and thus the form \refeq{eq:taueff3} is consistent with crystallography.}

The effective stress \refeq{eq:taueff3} together with the yield criterion \refeq{eq:yieldcrit} constitute the primary result of this paper. All stress components in \refeq{eq:taueff3} are invariant under the generalized inversion symmetry defined by \refeq{eq:symm2}. The presence of different auxiliary systems for $(\m,\n)$ (with the normal $\bm{n}'^\alpha$), and $(-\m,-\n)$ (with the normal $-\bm{n}''^\alpha$) is the consequence of choosing the angle of the auxiliary plane different from $90^\circ$. This projection cannot be used, because it would change $\sigma_{n'm}^\alpha$ to $-\sigma_{tm}^\alpha$, which is the stress that is not allowed by the symmetry conditions \refeq{eq:symm} and \refeq{eq:symm2}. \new{The chosen angle of the auxiliary plane ($60^\circ$) is only one possible choice and other orientations of the auxiliary plane would result in different forms of the effective stress. Nevertheless, all these effective stresses would be equally applicable in defining the yield criteria and predicting the onset of yielding.} Since no details have been given so far about the Miller indices representing the vectors $\m$ and $\n$, the effective stress \refeq{eq:taueff3} is applicable to all non-polar crystal structures. In the following, we will demonstrate its applicability to bcc and hcp crystals.


\section{Application to cubic and hexagonal crystals}

\subsection{Body-centered cubic crystals}

The plastic deformation of bcc metals is governed by the glide of $1/2\gdir{111}$ screw dislocations moving by elementary steps on $\gplane{110}$ planes. We will be concerned with the usual (engineering) case, where the initiation of plastic deformation is not affected by the hydrostatic stress and by the tension/compression acting parallel to the dislocation line. Under these assumptions, the effective stress \refeq{eq:taueff3} reduces to
\begin{equation}
  \tau^{*\alpha} = \sigma_{nm}^\alpha + a_1\sigma_{n'm}^\alpha + a_2(\sigma_{nn}^\alpha-\sigma_{tt}^\alpha) + a_3(\sigma_{n'n'}^\alpha-\sigma_{t't'}^\alpha) \ .
  \label{eq:taueff3_bcc}
\end{equation}
The individual stress components are expressed as explained in Section~\ref{sec:theory} with the (non-normalized) vectors $\bm{m}^\alpha$, $\bm{n}^\alpha$, and $\bm{m}'^\alpha$ summarized for the 24 slip systems in \reftab{tab:bcc24sys}. The coefficients $a_1$, $a_2$ and $a_3$ in \refeq{eq:taueff3_bcc} as well as the value of $\tau^*_{cr}$ in \refeq{eq:yieldcrit} need to be determined using atomistic simulations or from experiments. In the former case, the parameterization may follow the procedure described in \citet{groger:08a}. This is based on the observation that only the shear stresses perpendicular and parallel to the slip direction affect the CRSS to move $1/2\gdir{111}$ screw dislocations in bcc metals \citep{duesbery:84, ito:01, groger:14a}. The twinning-antitwinning asymmetry in bcc metals \new{\citep{christian:83}} dictates that the coefficient $a_1$ must be positive so that the CRSS to move the dislocation is higher when shearing the crystal in the antitwinning sense than when shearing it in the twinning sense. The magnitudes and signs of the coefficients $a_2$ and $a_3$ are determined by the dependence of the CRSS on shear stresses perpendicular to the slip direction. 

\begin{table}
  \caption{The 24 slip systems in bcc crystals. The crystallographic
  vectors $\bm{m}^\alpha$, $\bm{n}^\alpha$, $\bm{n}'^\alpha$ have to be normalized before their
  use in \reftab{tab:sigcomp} and evaluation of the effective stress \refeq{eq:taueff3_bcc}.
  \label{tab:bcc24sys}}
  \begin{tabular}{lccccclcccc}
    \hline    
    $\alpha$ & system & $\bm{m}^\alpha$ & $\bm{n}^\alpha$ & $\bm{n}'^\alpha$ &&
    $\alpha$ & system & $\bm{m}^\alpha$ & $\bm{n}^\alpha$ & $\bm{n}'^\alpha$ \\
    \cline{1-5} \cline{7-11}
    \systab {1} {$01\bar{1}$} {$111$} {$\bar{1}10$} &&
      \systab {$1^*$} {$01\bar{1}$} {$\bar{1}\bar{1}\bar{1}$} {$10\bar{1}$} \\
    \systab {2} {$\bar{1}01$} {$111$} {$0\bar{1}1$} &&
      \systab {$2^*$} {$\bar{1}01$} {$\bar{1}\bar{1}\bar{1}$} {$\bar{1}10$} \\
    \systab {3} {$1\bar{1}0$} {$111$} {$10\bar{1}$} &&
      \systab {$3^*$} {$1\bar{1}0$} {$\bar{1}\bar{1}\bar{1}$} {$0\bar{1}1$} \\
    \cline{1-5} \cline{7-11}
    \systab {4} {$\bar{1}0\bar{1}$} {$\bar{1}11$} {$\bar{1}\bar{1}0$} && 
      \systab {$4^*$} {$\bar{1}0\bar{1}$} {$1\bar{1}\bar{1}$} {$01\bar{1}$} \\
    \systab {5} {$0\bar{1}1$} {$\bar{1}11$} {$101$} && 
      \systab {$5^*$} {$0\bar{1}1$} {$1\bar{1}\bar{1}$} {$\bar{1}\bar{1}0$} \\
    \systab {6} {$110$} {$\bar{1}11$} {$01\bar{1}$} && 
      \systab {$6^*$} {$110$} {$1\bar{1}\bar{1}$} {$101$} \\
    \cline{1-5} \cline{7-11}
    \systab {7} {$0\bar{1}\bar{1}$} {$\bar{1}\bar{1}1$} {$1\bar{1}0$} && 
      \systab {$7^*$} {$0\bar{1}\bar{1}$} {$11\bar{1}$} {$\bar{1}0\bar{1}$} \\
    \systab {8} {$101$} {$\bar{1}\bar{1}1$} {$011$} && 
      \systab {$8^*$} {$101$} {$11\bar{1}$} {$1\bar{1}0$} \\
    \systab {9} {$\bar{1}10$} {$\bar{1}\bar{1}1$} {$\bar{1}0\bar{1}$} && 
      \systab {$9^*$} {$\bar{1}10$} {$11\bar{1}$} {$011$} \\
    \cline{1-5} \cline{7-11}
    \systab {10} {$10\bar{1}$} {$1\bar{1}1$} {$110$} && 
      \systab {$10^*$} {$10\bar{1}$} {$\bar{1}1\bar{1}$} {$0\bar{1}\bar{1}$} \\
    \systab {11} {$011$} {$1\bar{1}1$} {$\bar{1}01$} && 
      \systab{$11^*$} {$011$} {$\bar{1}1\bar{1}$} {$110$} \\
    \systab {12} {$\bar{1}\bar{1}0$} {$1\bar{1}1$} {$0\bar{1}\bar{1}$} && 
      \systab{$12^*$} {$\bar{1}\bar{1}0$} {$\bar{1}1\bar{1}$} {$\bar{1}01$} \\
    \hline    
  \end{tabular}
\end{table}

\begin{figure}
  \centering
  \includegraphics[scale=.6]{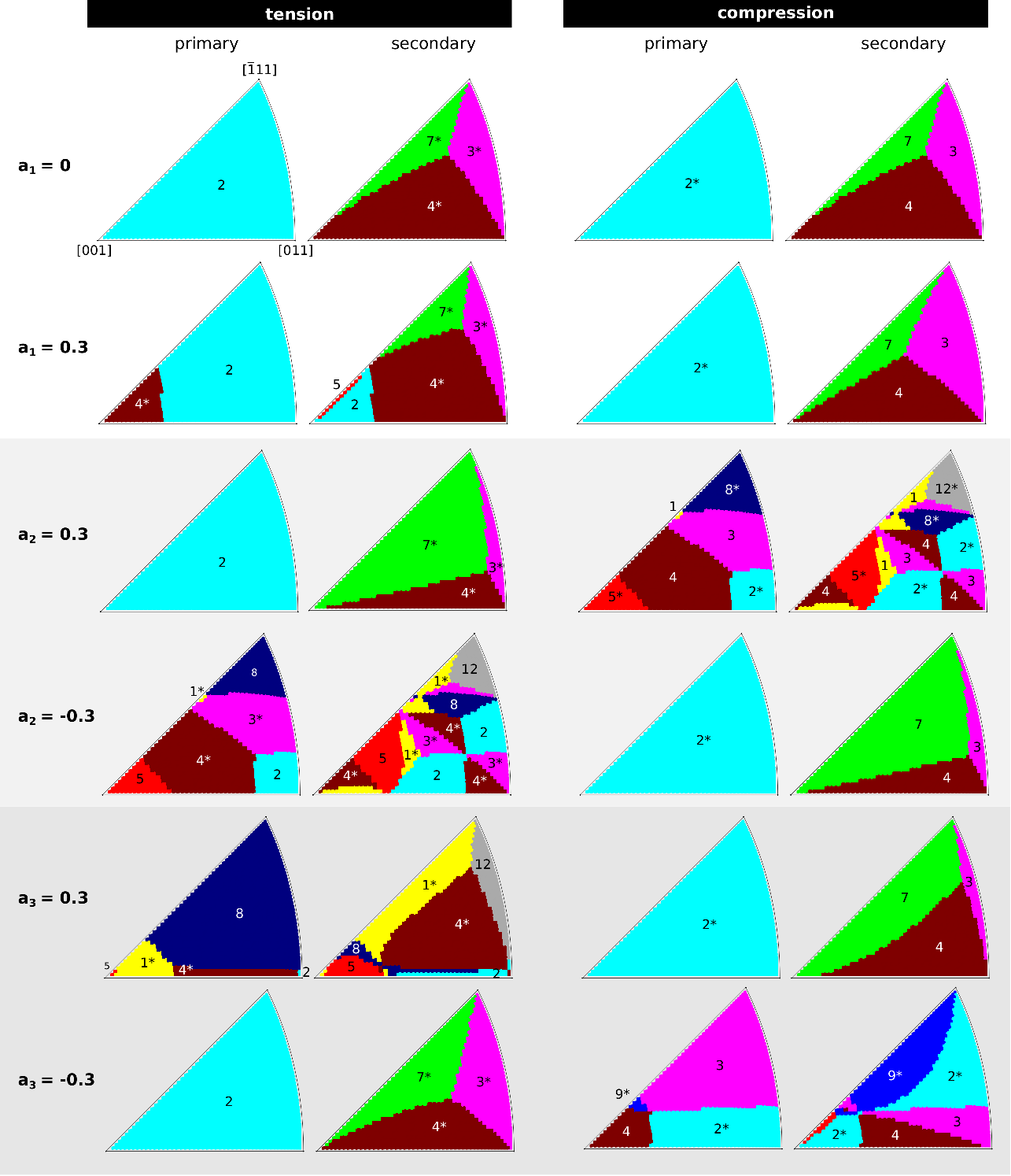}
  \caption{Predictions of the symmetry-adapted yield criterion for bcc metals \refeq{eq:taueff3_bcc} for different choices of the parameters $a_1$, $a_2$ and $a_3$ shown on the left (the remaining parameters are zero). The numbers refer to the slip systems in \reftab{tab:bcc24sys}.}
  \label{fig:slipact_bcc_tc}
\end{figure}

In \reffig{fig:slipact_bcc_tc}, we demonstrate how different choices of the coefficients $a_1$, $a_2$ and $a_3$, and thus the three non-Schmid stress components in \refeq{eq:taueff3_bcc}, affect the prediction of the primary and secondary slip systems (i.e., the first two systems $\alpha$ for which the value of $\tau^{*\alpha}$ reaches $\tau^*_{cr}$). These calculations were made for uniaxial tension and compression in all directions covering the area of the stereographic triangle $[001]-[011]-[\bar{1}11]$. Preliminary parameterizations of the yield criterion \refeq{eq:taueff3_bcc} for existing atomistic studies on bcc metals suggest that both coefficients $a_2$ and $a_3$ may be negative, whereas $a_1$ must always be non-negative to obtain correct trend of the twinning-antitwinning asymmetry. In the following, we thus consider $a_1=\{0,0.3\}$, and $a_2,a_3=\pm0.3$. The first row of images gives the predictions of the Schmid law that were obtained by setting $a_1=a_2=a_3=0$. The primary and secondary slip systems predicted for tension and compression are obviously identical, as dictated by the symmetry between tension and compression embodied in the Schmid law. The remaining maps demonstrate how individual non-Schmid stress components (i.e., nonzero values of $a_1$, $a_2$, or $a_3$) cause deviations of the slip activity from the prediction of the Schmid law. The effect of the twinning-antitwinning asymmetry is shown in the second row for $a_1=0.3$ (the remaining two parameters are zero). For most orientations, the primary slip system is still number 2 in tension (number 2* in compression). This generally agrees with atomistic simulations, where the application of pure shear stress parallel to the slip direction results in slip on the most highly stresses slip system 2 (2*). The effect of the shear stress perpendicular to the slip direction is demonstrated in the maps for nonzero values of the coefficients $a_2$ and $a_3$. The presence of these shear stresses in the yield criterion makes the predictions of the slip activity more orientational-dependent. For $a_2>0$ or $a_3<0$, loading in tension in the center-triangle orientation gives a similar prediction as the Schmid law, i.e. primary slip on the system 2 (2*). The same is true for compression when $a_2<0$ or $a_3>0$. However, very different predictions are obtained for $a_2<0$ or $a_3>0$ when loading in tension, as well as for $a_2>0$ or $a_3<0$ if the load is applied in compression. In these four cases, the primary slip is predicted to occur on the slip system 4 (4*), 3 (3*), or 8 (8*) for most orientations of the loading axis.

\subsection{Hexagonal crystals}

\begin{figure}
  \centering
  \includegraphics[scale=1.0]{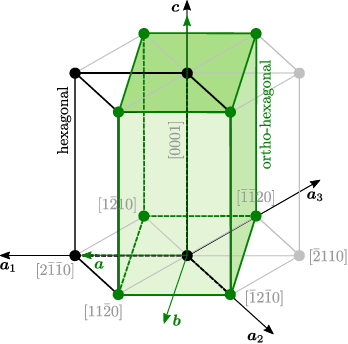}
  \caption{The primitive hexagonal cell (black) and the ortho-hexagonal cell (green) with the lattice vectors shown by the same color. Several low-index directions in the hexagonal lattice are shown in gray.}
  \label{fig:hex_lattice}
\end{figure}

In hexagonal crystallography, it is customary to characterize directions and normals of planes using four-index notations. The first three indices constitute projections onto the axes $\bm{a}_1$, $\bm{a}_2$, $\bm{a}_3$ in the basal plane, whereas the fourth is the projection onto the $\bm{c}$ axis, as shown in \reffig{fig:hex_lattice}. The addition of the third axis, i.e. $\bm{a}_3=-(\bm{a}_1+\bm{a}_2)$, ensures that crystallographically equivalent directions have similar Miller indices in the four-index notation. In order to simplify mathematical operations in the hexagonal system, \citet{otte:65} proposed to convert the four-index representations of slip directions and slip plane normals in the hexagonal system into an ortho-hexagonal system spanned by mutually orthogonal lattice vectors $\bm{a}$, $\bm{b}$, $\bm{c}$ shown in \reffig{fig:hex_lattice}. We will use these projection rules in the following to derive expressions for the six stress components $\sigma_{ij}$ in \refeq{eq:taueff3} directly in terms of the slip directions and slip plane normals defined in the four-index hexagonal notation. 

\subsubsection{Slip directions in the ortho-hexagonal basis}

Consider an arbitrary direction $\bm{m}$ that is represented by the Miller indices $[uvtw]$ in the four-index hexagonal system. The vector $\bm{m}$ can be recovered from these Miller indices using the basis vectors of the hexagonal system, i.e. $\bm{m} = u\bm{a}_1 + v\bm{a}_2 + (u+v)(\bm{a}_1+\bm{a}_2) + w\bm{c}$, where we have used well-known identities $t=-(u+v)$ and $\bm{a}_3=-(\bm{a}_1+\bm{a}_2)$. In the ortho-hexagonal system, the same direction $\bm{m}$ is represented by the Miller indices $[pqr]$, i.e. $\bm{m} = p\bm{a} + q\bm{b} + r\bm{c}$. According to \citet{otte:65} the relations between the lattice vectors in the two systems are $\bm{a} = \bm{a}_1$, $\bm{b} = \bm{a}_1 + 2\bm{a}_2$, whereas the third axis remains the same in both systems. Equating the right-hand sides of the two representations of the vector $\bm{m}$ above and implementing these transformation rules provides the indices of directions in the ortho-hexagonal system in terms of those in the hexagonal system: $p=3u/2$, $q=u/2+v$, $r=w$. It is clear that the third index in $[uvtw]$ is redundant and can be omitted (it is replaced in the following by a dot). Therefore, any direction represented by the vector $\bm{M}=[uv.w]$ defined in the hexagonal system can be expressed by the vector $\bm{m}=[pqr]$ in the ortho-hexagonal system as
\begin{equation}
  \bm{m} = \bm{M}\bm{Q} \ ,
  \label{eq:mfromM}
\end{equation}
where both $\bm{m}$ and $\bm{M}$ are row vectors with three values. The transformation matrix $\bm{Q}$ has the form
\begin{equation}
  \bm{Q} = \left[
    \begin{array}{ccc}
      3/2 & 1/2 & 0  \\
      0 & 1 & 0 \\
      0 & 0 & 1
    \end{array}
    \right] \ .
  \label{eq:Q}
\end{equation}

It is important to emphasize that the indices $[pqr]$ representing $\bm{m}$ in the ortho-hexagonal basis refer to the basis vectors  $\bm{a}=a\hat{\bm{a}}$, $\bm{b}=a\sqrt{3}\hat{\bm{b}}$ and $\bm{c}=c\hat{\bm{c}}$,
where $\hat{\bm{a}}$, $\hat{\bm{b}}$, $\hat{\bm{c}}$ are unit vectors along the three directions. In terms of these basis vectors, $\bm{m}=[pqr]$ can be expressed as
\begin{equation}
  \bm{m} = pa\hat{\bm{a}} + qa\sqrt{3}\hat{\bm{b}} + rc\hat{\bm{c}} \ .
\end{equation}
We want to find the unit vector $\hat{\bm{m}}=[\hat{p}\hat{q}\hat{r}]$ parallel to $\bm{m}$, which we define as
\begin{equation}
  \hat{\bm{m}} = \hat{p}\hat{\bm{a}} + \hat{q}\hat{\bm{b}} + \hat{r}\hat{\bm{c}} \ .
\end{equation}
The two vectors are related by $\hat{\bm{m}} = \bm{m}/\lambda_m$, where $\lambda_m$ is a normalization factor. This condition gives three equations that define the components of the unit vector $\hat{\bm{m}}$: $\hat{p} = pa/\lambda_m$, $\hat{q} = qa\sqrt{3}/\lambda_m$, $\hat{r} = rc/\lambda_m$. Substituting these equations into the normalization condition $\hat{p}^2 + \hat{q}^2 + \hat{r}^2 = 1$ provides the normalization factor $\lambda_m = a\sqrt{p^2 + 3q^2 + (c/a)^2r^2}$. It is evident that substituting $\lambda_m$ back into the expressions for $\hat{p}$,  $\hat{q}$, $\hat{r}$ eliminates the lattice parameter $a$. For the sake of simplicity, we thus redefine the components of the unit vector $\hat{\bm{m}}=[\hat{p}\hat{q}\hat{r}]$ and the normalization factor as 
\begin{eqnarray}
  &\hat{p} = p/\lambda_m \ , \quad
  \hat{q} = q\sqrt{3}/\lambda_m \ , \quad
  \hat{r} = r(c/a)/\lambda_m& \\ \nonumber
  &\lambda_m = \sqrt{p^2 + 3q^2 + (c/a)^2r^2}& \ .
\end{eqnarray}
Employing \refeq{eq:mfromM} in the definition of $\hat{\bm{m}}$ allows to express the unit vector parallel to the slip direction in the (orthonormal) ortho-hexagonal basis simply as a division by the factor $\lambda_m$:
\begin{equation}
  \hat{\bm{m}} = \frac{\bm{M} \bm{Q}}{\lambda_m} \ .
  \label{eq:munit}
\end{equation}


\subsubsection{Slip plane normals in the ortho-hexagonal basis}

Different transformation rules apply for directions of plane normals \citep{otte:65}. To demonstrate this, consider an arbitrary plane defined by the indices $(hkil)$ in the four-index hexagonal system, where $i=-(h+k)$ as explained above. The vector normal to this plane is then $\bm{n} = h\bm{a}_1^* + k\bm{a}_2^* + l\bm{c}^*$, where $\bm{a}_1^*, \bm{a}_2^*, \bm{c}^*$ are reciprocal lattice vectors. In the ortho-hexagonal system, this plane is represented by the Miller index $(efg)$, which defines the same vector $\bm{n} = e\bm{a}^* + f\bm{b}^* + g\bm{c}^*$. The transformation rules between the two reciprocal bases are $\bm{a}^* = \bm{a}_1^*-\bm{a}_2^*/2$, $\bm{b}^* = \bm{a}_2^*/2$ with $\bm{c}^*$ being parallel in both bases. Equating the factors in front of $\bm{a}_1^*$, $\bm{a}_2^*$ and $\bm{c}^*$ provides the relations between the indices in the three-index ortho-hexagonal system in terms of the indices in the four-index hexagonal system: $e=h$, $f=h+2k$, $g=l$. Similarly as above, the third index in $(hkil)$ is redundant and does not enter the transformation (it is again replaced by a dot). Therefore, the vector normal to the plane $(hk.l)$ in the hexagonal system, represented by the vector $\bm{N}=[hk.l]$, can be expressed as the vector $\bm{n}=[efg]$ in the ortho-hexagonal system as
\begin{equation}
  \bm{n} = \bm{N}\bm{Q}^* \ ,
  \label{eq:nfromN}
\end{equation}
where both $\bm{n}$ and $\bm{N}$ are again row vectors with three values. The transformation matrix $\bm{Q}^*$ has the form
\begin{equation}
  \bm{Q}^* = \left[
    \begin{array}{ccc}
      1 & 1 & 0 \\
      0 & 2 & 0 \\
      0 & 0 & 1
    \end{array}
    \right] \ .
  \label{eq:Qstar}
\end{equation}

Similar procedure as above can be applied to derive the normalization factors for plane normals. In the ortho-hexagonal basis, the basis vectors for plane normals are $\bm{a}^* = \bm{a}/a^2$, $\bm{b}^* = \bm{b}/3a^2$, $\bm{c}^* = \bm{c}/c^2$. Using these basis vectors, the plane normals $\bm{n}=[efg]$ can be expressed as 
\begin{equation}
  \bm{n} = \frac{e}{a}\hat{\bm{a}} + \frac{f}{a\sqrt{3}}\hat{\bm{b}} + \frac{g}{c}\hat{\bm{c}} \ .
\end{equation}
We want to find the unit vector $\hat{\bm{n}}=[\hat{e}\hat{f}\hat{g}]$ parallel to $\bm{n}$, which we define as
\begin{equation}
  \hat{\bm{n}} = \hat{e}\hat{\bm{a}} + \hat{f}\hat{\bm{b}} + \hat{g}\hat{\bm{c}} \ .
\end{equation}
The two vectors are related by $\hat{\bm{n}} = \bm{n}/\lambda_n$, where $\lambda_n$ is a normalization factor. This condition gives three equations that define the components of the unit vector $\hat{\bm{n}}$: $\hat{e} = e/a\lambda_n$, $\hat{f} = f/a\sqrt{3}\lambda_n$, $\hat{g} = g/c\lambda_n$. Substituting these equations into the normalization condition $\hat{e}^2 + \hat{f}^2 + \hat{g}^2 = 1$ provides the normalization factor $\lambda_n = (1/a)\sqrt{e^2 + f^2/3 + g^2/(c/a)^2}$. It is again evident that substituting $\lambda_n$ back into the expressions for $\hat{e}$, $\hat{f}$, $\hat{g}$ eliminates the lattice parameter $a$. We thus redefine the components of the unit vector  $\hat{\bm{n}}=[\hat{e}\hat{f}\hat{g}]$ and the normalization factor as
\begin{eqnarray}
  &\hat{e} = e/\lambda_n \ , \quad
  \hat{f} = f/\sqrt{3}\lambda_n \ , \quad
  \hat{g} = g/(c/a)\lambda_n& \\ \nonumber
  &\lambda_n = \sqrt{e^2 + f^2/3 + g^2/(c/a)^2}& \ .
\end{eqnarray}
Employing \refeq{eq:nfromN} in the definition of $\hat{\bm{n}}$ allows to express the  unit normal of the slip plane in the (orthonormal) ortho-hexagonal basis simply as a division by the factor $\lambda_n$:
\begin{equation}
  \hat{\bm{n}} = \frac{\bm{N} \bm{Q}^*}{\lambda_n} \ .
  \label{eq:nunit}
\end{equation}


\subsubsection{Slip activity in hexagonal crystals}

The stress components in \refeq{eq:taueff3} were defined using the slip directions and slip plane normals expressed in two orthogonal coordinate systems $(\t,\n,\m)$ and $(\bm{t}'^\alpha,\bm{n}'^\alpha,\m)$ misoriented by $60^\circ$. Using the linear transformations \refeq{eq:munit} and \refeq{eq:nunit}, we can now express these vectors in terms of the vectors $\MM$, $\NN$, $\NNN$ in \reftab{tab:hexsys} that are natural to the hexagonal system as $\mm = (1/\lambda_m)\bm{M}^\alpha\bm{Q}$, $\nn = (1/\lambda_n)\bm{N}^\alpha\bm{Q}^*$ and $\nnn = (1/\lambda_{n'})\bm{N}'^\alpha\bm{Q}^*$. Using these relations, we determine the six stress components entering \refeq{eq:taueff3} directly using the Miller indices of slip planes and slip directions given in \reftab{tab:hexsys}. To demonstrate this, consider the Schmid stress that was previously defined as $\sigma_{nm}^\alpha = \nn\bm{\sigma}(\mm)^T $. Employing the relations above provides an equivalent expression, $\sigma_{nm}^\alpha = (1/\lambda_n\lambda_m)\NN\bm{\Sigma}^*(\MM)^T$, where $\bm{\Sigma}^*= \bm{Q}^*\bm{\sigma}\bm{Q}^T$ is a $3\times 3$ matrix that can be regarded as a representation of the stress tensor $\bm{\sigma}$ in the hexagonal system. It is interesting to observe that the previous two expressions of $\sigma_{nm}^\alpha$ have the same mathematical form and differ only by the stress tensor in the two bases and by the normalization factor. The same is true also for the stress components $\sigma_{n'm}^\alpha$, $\sigma_{nn}^\alpha$, $\sigma_{n'n'}^\alpha$, and $\sigma_{mm}^\alpha$ as shown in \reftab{tab:sigcomp} (column ``hcp crystals'').

%
%
\begin{table}[!htb]
  \centering \footnotesize
  \caption{Hexagonal slip systems with the slip directions $\MM$, slip plane normals $\NN$ and the normals of auxiliary planes $\NNN$ given in the four-index (hexagonal) basis. Separate numbering is used here for the three types of slip systems.}
  \label{tab:hexsys}
  \begin{tabular}{clccccc}
    \hline
    type & $\alpha$ & system & $\bm{M}^\alpha$ & $\bm{N}^\alpha$ & $\bm{N}'^\alpha$ \\
    \hline
          & 1  & $(0001)[11\bar{2}0]$       & $[11\bar{2}0]$ & $[0001]$ & $\new{[1,-1,0,\sqrt{2/3}]}$ \\
          & 2  & $(0001)[1\bar{2}10]$       & $[1\bar{2}10]$ & $[0001]$ & $\new{[-1,0,1,\sqrt{2/3}]}$ \\
    basal & 3  & $(0001)[\bar{2}110]$       & $[\bar{2}110]$ & $[0001]$ & $\new{[0,1,-1,2/\sqrt{3}]}$ \\
          \cline{2-6}
          & $1^*$ & $(0001)[\bar{1}\bar{1}20]$       & $[\bar{1}\bar{1}20]$ & $[0001]$ & $\new{[-1,1,0,\sqrt{2/3}]}$ \\
          & $2^*$ & $(0001)[\bar{1}2\bar{1}0]$       & $[\bar{1}2\bar{1}0]$ & $[0001]$ & $\new{[1,0,-1,\sqrt{2/3}]}$ \\
          & $3^*$ & $(0001)[2\bar{1}\bar{1}0]$       & $[2\bar{1}\bar{1}0]$ & $[0001]$ & $\new{[0,-1,1,2/\sqrt{3}]}$ \\
    \hline
              & 4  & $(\bar{1}100)[11\bar{2}0]$ & $[11\bar{2}0]$ & $[\bar{1}100]$ & $\new{[-1,1,0,\sqrt{6}]}$ \\
              & 5  & $(\bar{1}010)[1\bar{2}10]$ & $[1\bar{2}10]$ & $[\bar{1}010]$ & $\new{[-1,0,1,-\sqrt{6}]}$ \\
    prismatic & 6  & $(0\bar{1}10)[\bar{2}110]$ & $[\bar{2}110]$ & $[0\bar{1}10]$ & $\new{[0,-1,1,2\sqrt{3}]}$ \\
              \cline{2-6}
              & $4^*$ & $(\bar{1}100)[\bar{1}\bar{1}20]$ & $[\bar{1}\bar{1}20]$ & $[\bar{1}100]$ & $\new{[-1,1,0,-\sqrt{6}]}$ \\
              & $5^*$ & $(\bar{1}010)[\bar{1}2\bar{1}0]$ & $[\bar{1}2\bar{1}0]$ & $[\bar{1}010]$ & $\new{[-1,0,1,\sqrt{6}]}$ \\
              & $6^*$ & $(0\bar{1}10)[2\bar{1}\bar{1}0]$ & $[2\bar{1}\bar{1}0]$ & $[0\bar{1}10]$ & $\new{[0,-1,1,-2\sqrt{3}]}$ \\
    \hline
              & 1  & $(\bar{1}011)[11\bar{2}3]$ & $[11\bar{2}3]$ & $[\bar{1}011]$ & $\new{[-1+3/\sqrt{2},-3/\sqrt{2},1,1]}$ \\
              & 2  & $(0\bar{1}11)[11\bar{2}3]$ & $[11\bar{2}3]$ & $[0\bar{1}11]$ & $\new{[5+5/\sqrt{2},-4-5/\sqrt{2},-1,-1]}$ \\
              & 3  & $(\bar{1}101)[1\bar{2}13]$ & $[1\bar{2}13]$ & $[\bar{1}101]$ & $\new{[-1-3/\sqrt{2},1,3/\sqrt{2},1]}$ \\
              & 4  & $(01\bar{1}1)[1\bar{2}13]$ & $[1\bar{2}13]$ & $[01\bar{1}1]$ & $\new{[-5+5/\sqrt{2},1,4-5/\sqrt{2},1]}$ \\
              & 5  & $(1\bar{1}01)[\bar{2}113]$ & $[\bar{2}113]$ & $[1\bar{1}01]$ & $\new{[1,(7-3\sqrt{6})/2,3(-3+\sqrt{6})/2,1]}$ \\
    pyramidal & 6  & $(10\bar{1}1)[\bar{2}113]$ & $[\bar{2}113]$ & $[10\bar{1}1]$ & $\new{[-1,3(3+\sqrt{6})/2,1-3(3+\sqrt{6})/2,-1]}$ \\
              \cline{2-6}
              & $1^*$ & $(\bar{1}011)[\bar{1}\bar{1}2\bar{3}]$ & $[\bar{1}\bar{1}2\bar{3}]$ & $[\bar{1}011]$ & $\new{[-1-3/\sqrt{2},3/\sqrt{2},1,1]}$ \\
              & $2^*$ & $(0\bar{1}11)[\bar{1}\bar{1}2\bar{3}]$ & $[\bar{1}\bar{1}2\bar{3}]$ & $[0\bar{1}11]$ & $\new{[-5+5/\sqrt{2},4-5/\sqrt{2},1,1]}$ \\
              & $3^*$ & $(\bar{1}101)[\bar{1}2\bar{1}\bar{3}]$ & $[\bar{1}2\bar{1}\bar{3}]$ & $[\bar{1}101]$ & $\new{[-1+3/\sqrt{2},1,-3/\sqrt{2},1]}$ \\
              & $4^*$ & $(01\bar{1}1)[\bar{1}2\bar{1}\bar{3}]$ & $[\bar{1}2\bar{1}\bar{3}]$ & $[01\bar{1}1]$ & $\new{[5+5/\sqrt{2},-1,-4-5/\sqrt{2},-1]}$ \\
              & $5^*$ & $(1\bar{1}01)[2\bar{1}\bar{1}\bar{3}]$ & $[2\bar{1}\bar{1}\bar{3}]$ & $[1\bar{1}01]$ & $\new{[-1,-(7+3\sqrt{6})/2,3(3+\sqrt{6})/2,-1]}$ \\
              & $6^*$ & $(10\bar{1}1)[2\bar{1}\bar{1}\bar{3}]$ & $[2\bar{1}\bar{1}\bar{3}]$ & $[10\bar{1}1]$ & $\new{[1,3(-3+\sqrt{6})/2,(7-3\sqrt{6})/2,1]}$ \\
    \hline
  \end{tabular}
\end{table}

The expressions for $\sigma_{tt}^\alpha$ and $\sigma_{t't'}^\alpha$ in the hexagonal system are quite different from those in cubic systems. By definition, $\sigma_{tt}^\alpha = (\nn\times\mm)\bm{\sigma}(\nn\times\mm)^T$, where the slip direction $\mm$ and the slip plane normal $\nn$ are expressed in terms of $\MM$ and $\NN$ as explained above. This substitution gives
\begin{equation}
  \sigma_{tt}^\alpha = \frac{[(\bm{N}^\alpha\bm{Q}^*)\times(\bm{M}^\alpha\bm{Q})]\bm{\sigma}[(\bm{N}^\alpha\bm{Q}^*)\times(\bm{M}^\alpha\bm{Q})]^T}{\lambda_n^2\lambda_m^2} \ .
  \label{eq:sigtt_apx}
\end{equation}
We want to derive an equivalent form of this expression in which the vectors $\bm{N}^\alpha$ and $\bm{M}^\alpha$ are decoupled from the multiplication with the matrices $\bm{Q}^*$ and $\bm{Q}$, respectively. For that purpose, it is first convenient to write the cross-products in \refeq{eq:sigtt_apx} in tensorial form as $[(\bm{N}^\alpha\bm{Q}^*)\times(\bm{M}^\alpha\bm{Q})]_k = (\epsilon_{i'j'k}Q^*_{ii'}Q_{jj'})N_{i}M_{j}$. Incorporating this back into \refeq{eq:sigtt_apx} provides an equivalent expression for this stress component,
\begin{equation}
  \sigma_{tt}^\alpha = \frac{(\bm{N}^\alpha\otimes\bm{M}^\alpha)_{ij} \Sigma_{ijpq} (\bm{N}^\alpha\otimes\bm{M}^\alpha)_{pq}}{\lambda_n^2\lambda_m^2} \ ,
  \label{eq:sigtt_final_apx}
\end{equation}
where $\NN\otimes\MM=N_i^\alpha M_j^\alpha$, and $\Sigma_{ijpq} = (\epsilon_{i'j'k}Q^*_{ii'}Q_{jj'})\sigma_{kl}(\epsilon_{lp'q'}Q^*_{pp'}Q_{qq'})$ is a representation of the stress tensor in the hexagonal basis. The two terms in the brackets of $\Sigma_{ijpq}$ are cross-products with respect to the second indices of the matrices $\bm{Q}$ and $\bm{Q}^*$. Unlike all other stress components, the matrix form of $\sigma_{tt}^\alpha$ in hexagonal crystals \refeq{eq:sigtt_final_apx} is different from its expression for cubic systems. This is due to the cross-product between the slip plane normal and the slip direction, which transform differently between the ortho-hexagonal and hexagonal systems as shown in \refeq{eq:mfromM} and \refeq{eq:nfromN}. The same argument applies to the stress component $\sigma_{t't'}^\alpha$ that is obtained from \refeq{eq:sigtt_final_apx} by replacing $\bm{N}^\alpha$ with $\bm{N}'^\alpha$ and $\lambda_n$ with $\lambda_{n'}$.

\begin{figure}
  \centering
  \includegraphics[scale=.6]{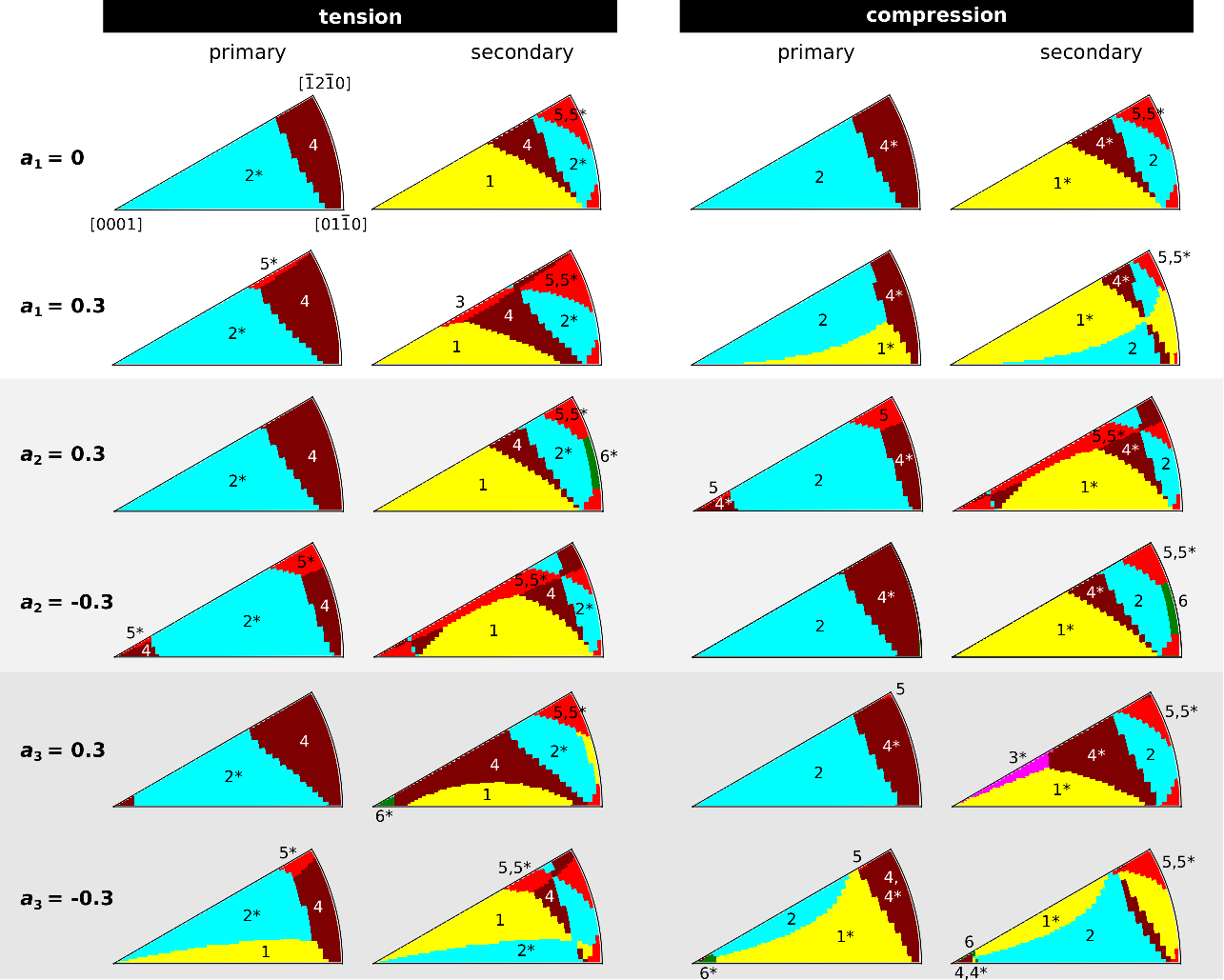}
  \caption{Predictions of the symmetry-adapted yield criterion for hexagonal metals deforming by basal-prismatic slip for different choices of the parameters $a_1$, $a_2$ and $a_3$ shown on the left (the remaining parameters are zero). The numbers refer to the slip systems in \reftab{tab:hexsys}.}
  \label{fig:slipact_hex_tc}
\end{figure}

Similarly as in the case of bcc metals, it is instructive to look at the predictions of this yield criterion for hexagonal metals, which is defined by \refeq{eq:taueff3} with the individual stresses given in \reftab{tab:sigcomp} (column ``hcp crystals''). For simplicity, we again ignore the last two components in \refeq{eq:taueff3}, because they do not affect the onset of plastic deformation at low strain rates. We particularly focus on $1/3\gdir{11\bar{2}0}$ dislocations, which may glide on both basal and prismatic planes. In the following, we thus consider only the twelve systems 1-3, 1*-3*, 4-6, 4*-6* in \reftab{tab:hexsys}. The stereographic triangle in which the analysis of operative slip systems will be made is now  $[0001]-[01\bar{1}0]-[\bar{1}2\bar{1}0]$. Due to lower symmetry of hexagonal crystals, this triangle contains three elementary stereographic triangles considered in \reffig{fig:slipact_bcc_tc} for crystals with cubic symmetry. The first row in \reffig{fig:slipact_hex_tc} again shows the prediction of the Schmid law ($a_1=a_2=a_3=0$), where the symmetry between tension and compression is evident. Loading in the center of the stereographic triangle results in primary basal slip on the system 2 (2*) and secondary basal slip on the system 1 (1*). However, deviation of the loading axis towards the $[01\bar{1}0]-[\bar{1}2\bar{1}0]$ edge of the triangle activates primary prismatic slip on the system 4 (4*) with the secondary contribution arising from basal or prismatic slip, depending on the orientation. Similarly as in bcc metals, the effects of individual non-Schmid stress components depend on the nature of the particular stress term and on the character of the applied load. For some combinations of $a_1$, $a_2$, $a_3$, these non-Schmid stresses result in shifting of the area of basal slip on the system 2 (2*) and/or 1 (1*), but other combinations promote prismatic slip on the systems 4 (4*) and/or 5 (5*). The precise values of the coefficients $a_1$, $a_2$ and $a_3$, and thus predictions of the slip activity for a particular hexagonal metal, have to be again obtained from atomistic simulations or from experiments.

\begin{table}
  \caption{Table of explicit representations of the six components of the stress tensor that appear in the effective yield criteria for bcc and hcp crystals derived in this paper. The hydrostatic stress is $\sigma_h = (\sigma_{tt}+\sigma_{nn}+\sigma_{mm})/3$. The generalized stress tensors are $\bm{\Sigma} = \bm{Q}\bm{\sigma}\bm{Q}^T$, $\bm{\Sigma}^* = \bm{Q}^*\bm{\sigma}\bm{Q}^T$, $\bm{\Sigma}^{**} = \bm{Q}^*\bm{\sigma}\bm{Q}^{*T}$, and the definition of the tensor $\Sigma_{ijpq}$ is given below Eq.~\refeq{eq:sigtt_final_apx}. Notations: $\bm{A}\otimes \bm{B} = A_iB_j = C_{ij}$ and $\bm{\sigma}:\bm{C} = \sigma_{ij}C_{ij}=$ scalar.}
  \label{tab:sigcomp}
  \begin{tabular}{ccc}
    \hline
    stress & bcc crystals & hcp crystals\\
    \hline
    $\sigma_{nm}^\alpha$   & $\bm{\sigma}:(\nn\otimes\mm)$ & $(1/\lambda_n\lambda_m) [\bm{\Sigma}^*:(\NN\otimes\MM)]$ \\
    $\sigma_{n'm}^\alpha$  & $\bm{\sigma}:(\nnn\otimes\mm)$ & $(1/\lambda_{n'}\lambda_m) [\bm{\Sigma}^*:(\NNN\otimes\MM)]$ \\
    $\sigma_{nn}^\alpha$   & $\bm{\sigma}:(\nn\otimes\nn)$ & $(1/\lambda_n^2) [\bm{\Sigma}^{**}:(\NN\otimes\NN)]$ \\
    $\sigma_{n'n'}^\alpha$ & $\bm{\sigma}:(\nnn\otimes\nnn)$ & $(1/\lambda_{n'}^2)[\bm{\Sigma}^{**}:(\NNN\otimes\NNN)]$ \\
    $\sigma_{mm}$         & $\bm{\sigma}:(\mm\otimes\mm)$ & $(1/\lambda_m^2) [\bm{\Sigma}:(\MM\otimes\MM)]$ \\
    $\sigma_{tt}^\alpha$   & $\bm{\sigma}:[(\nn\times\mm) \otimes (\nn\times\mm)]$ & $(1/\lambda_n^2\lambda_m^2) [(\NN\otimes\MM)_{ij} \Sigma_{ijpq} (\NN\otimes\MM)_{pq}]$ \\
      $\sigma_{t't'}^\alpha$ & $\bm{\sigma}:[(\nnn\times\mm) \otimes (\nnn\times\mm)]$ & $(1/\lambda_{n'}^2\lambda_m^2) [(\NNN\otimes\MM)_{ij} \Sigma_{ijpq} (\NNN\otimes\MM)_{pq}]$ \\
    \hline
  \end{tabular}
\end{table}


\section{Conclusions}

We have applied Neumann's principle to demonstrate that the single-crystal yield criteria for materials with non-polar structures must be invariant under a simultaneous change of sign of the slip direction ($\m\rightarrow -\m$) and the slip plane normal ($\n\rightarrow -\n$) for all slip systems $\alpha$ determined by the underlying space group. This symmetry is satisfied by the criterion proposed by \citet{tresca:1864} but it is violated in all recent generalizations of this model that aim to describe the onset of yielding in materials that do not obey the Schmid law. In the latter models, only certain combinations of the pairs $(\m,\n)$ may thus enter the yield criteria, whereas the conjugates $(-\m,-\n)$ are not allowed despite the fact that they are equivalent by symmetry.

To remove this drawback \new{and avoid ambiguity in interpreting the activity of individual slip systems}, we have explored the possibility to derive a general form of the effective stress that satisfies this symmetry. The effective stress is written as a sum of power series for each independent component of the stress tensor. For simplicity, we retain only the lowest order stress terms that are invariant under the symmetry above. This results in a homogeneous effective stress of order two in which the quadratic stresses are associated with shear stresses parallel and perpendicular to the slip direction. We show that the presence of these quadratic terms can be avoided by introducing an auxiliary system in the zone of the slip direction, which leads to generalization of the symmetry condition that involves also the normal of the auxiliary plane. This transformation provides a linear yield criterion that is conceptually similar to that proposed by \citet{qin:92} but, additionally, satisfies Neumann's principle. 

The obtained yield criterion can be used to describe the slip activity in bcc and hcp crystals as well as in other non-polar crystal structures. In the case of hexagonal symmetry, we use the transformation rules between ortho-hexagonal and hexagonal systems to express all stress terms using four-index representations of slip directions and slip plane normals. These expressions are very similar to those used for bcc lattices provided the applied stress tensor is also generalized into the four-index notation. Somewhat different forms are obtained for the stresses $\sigma_{tt}^\alpha$ and $\sigma_{t't'}^\alpha$, which follows from different transformation rules for slip directions and slip plane normals.

If non-Schmid stresses play no role in the initiation of plastic deformation, as it is the case in fcc metals and for basal slip in hcp metals, the effective stress derived here reduces to the Schmid stress. The theoretical framework established in this paper then reduces to the well-established maximum shear stress theory developed by \citet{tresca:1864}. 


\section*{Acknowledgments}

The author acknowledges stimulating discussions on this topic with Vaclav Vitek and John Bassani. This research was supported by the Czech Science Foundation, grant no.~19-23411S. It was carried out under the project CEITEC 2020 (LQ1601) with financial support from the Ministry of Education, Youth and Sports of the Czech Republic under the National Sustainability Programme II.


\bibliography{bibliography}

\end{document}